\documentclass{emulateapj}

\usepackage{graphicx} 
\usepackage{color} 
 
\newcommand{\be}{  \begin{eqnarray} }
\newcommand{\ee}{  \end{eqnarray} }

\def\taucr{\tau_{_{\rm CR}}}
\def\lambdacr{\lambda_{_{\rm CR}}}
\def\Pcr{P_{_{\rm CR}}}
\def\Fcr{F_{_{\rm CR}}}
\def\Lcr{L_{_{\rm CR}}}

\def\spose#1{\hbox to 0pt{#1\hss}}
\def\lta{\mathrel{\spose{\lower 3pt\hbox{$\mathchar"218$}}
     \raise 2.0pt\hbox{$\mathchar"13C$}}}
\def\gta{\mathrel{\spose{\lower 3pt\hbox{$\mathchar"218$}}
     \raise 2.0pt\hbox{$\mathchar"13E$}}}

\begin{document}

\shorttitle{Cosmic Ray Feedback}
\title{The Eddington Limit in Cosmic Rays:  An
Explanation for the Observed Faintness of Starbursting Galaxies}
\author{Aristotle Socrates\altaffilmark{1,2}, Shane W. Davis
\altaffilmark{3,4} and Enrico Ramirez-Ruiz\altaffilmark{3,4}}
\altaffiltext{1}{Department of Astrophysical Sciences, Princeton 
University, Peyton Hall-Ivy Lane, Princeton, NJ 08544; 
socrates@astro.princeton.edu}
\altaffiltext{2}{Hubble Fellow}
\altaffiltext{3}{Institute for
Advanced Study, Einstein Drive, Princeton, NJ 08540: swd@ias.edu;
enrico@ias.edu}
\altaffiltext{4}{Chandra Fellow}
\begin{abstract}
We show that the luminosity of a star forming galaxy is 
capped by the production and subsequent expulsion of cosmic 
rays from its interstellar medium.  By defining an Eddington
luminosity in cosmic rays, we show that the star formation 
rate of a given galaxy is limited by its mass content and the 
cosmic ray mean free path.  When the cosmic ray luminosity and 
pressure reaches a critical value as a result of vigorous 
star formation, hydrostatic balance is lost, a galactic
scale cosmic ray-driven
wind develops, and star formation is choked off.  Cosmic ray 
pressure-driven winds are likely to produce 
wind velocities in proportion to and significantly in excess 
of the galactic escape velocity.  

It is possible that cosmic ray feedback results in   
the Faber-Jackson relation for a plausible set of input parameters
that describe cosmic ray production and transport, which 
are calibrated by observations of the Milky Way's interstellar cosmic 
rays as well as other galaxies.
\end{abstract}
\section{Introduction}

In the context of galaxy formation, ``feedback'' processes
are commonly invoked to limit the formation of stars.  
Feedback can be ``intrinsic'' or ``extrinsic,'' where 
an intrinsic feedback mechanism regulates star formation 
directly at the molecular cloud level and an extrinsic 
mechanism controls the supply of stellar fuel on galactic
scales.     
 
There is long-standing observational evidence that extrinsic feedback
mechanisms operate.  For example, the luminosity of local galaxies
correlates with the stellar velocity dispersion, and thus
gravitational binding energy, of the entire galaxy itself (Faber \&
Jackson 1976).  Two popular extrinsic feedback mechanism are
thermally-driven winds powered by core-collapse SNe (see e.g.,
Chevalier \& Clegg 1985; Leitherer et al. 1999) and momentum-driven
winds powered by starburst radiation (Scoville 2003, hereafter SCO;
Murray et al. 2005, hereafter MQT).  In this work, we consider another
possibility: galactic winds driven by cosmic rays.

\subsection{Basic Idea:  The Eddington Limit in Cosmic Rays }

There is a theoretical upper-limit for the luminosity of 
gravitationally bound systems, the Eddington limit $L_{\rm edd}$,
given by
\be
L_{\rm edd}=\frac{4\pi GM_{\rm enc}m_pc }{\sigma_T}
\ee
where $M_{\rm enc}$ is the enclosed mass and $\sigma_T$ is the 
Thomson cross section.  Quasars, the most steadily luminous objects 
in the Universe, are thought to be powered by accretion onto a
black hole and thus their maximum luminosity is given by
\be
L^{\rm Q}_{\rm edd}\simeq 10^{47}M_9{\rm erg\,s^{-1}}
\label{quasarmax}
\ee   
where $M_9$ is the quasar black hole mass in units of 
$10^9M_{\odot}$.  In the case of elliptical galaxies, the mass of 
the black hole's host galaxy is $\sim 10^2 -10^3$ times that of the 
black hole (Magorrian et al. 1998).  However, rather than being 
the brightest persistent sources in the Universe, typical 
luminosities of bright star forming galaxies range from $~ 10^{45}
-10^{46}\,{\rm erg\, s^{-1}}$ -- at least a factor of ten
less powerful than their quasar counterparts.  

Photons are not the only type of radiation produced from
the act of star formation.  For every core-collapse supernova, 
a fraction $\epsilon\sim 0.1$ of the released energy $\sim 10^{50}
{\rm erg}$ is converted into cosmic rays (CRs).  Unlike photons, cosmic 
rays interact many times with the magnetized interstellar medium, and exchange
momentum with it, through pitch-angle scatterings with Alfv\`enic 
disturbances.  Though the energy injection rate of CRs is insignificant 
in comparison to other by-products of star formation such as photons,
winds, and supernova shocks, the rate at which they inject momentum 
is not. 

In the case of the Milky Way, the CR diffusion time $t_{\rm 
diff,CR} \sim 10^7{\rm yrs}$ over a characteristic scale of $H\sim 
1{\rm kpc}$.  Thus the characteristic mean free path for a Galactic
CR is given by
\be
\lambdacr\sim \frac{3H^2}{c\,t_{\rm diff,CR}}\sim 
1\,{\rm pc},  
\ee 
for 1-10 GeV CRs.      
Whereas the Thomson mean free path 
$\lambda_T\sim 10^{24}{\rm cm}$. 
Thus, CRs are much more highly coupled to the Galactic 
matter in comparison to photons.

Due to the higher level coupling, the threshold for the maximum 
allowable CR flux and luminosity is lower in comparison to 
the electron scattering case.  A rough theoretical upper limit
for the CR luminosity $L_{\rm edd, CR}$ is given by
\be
L_{\rm edd, CR}\simeq L_{\rm edd}\frac{\lambdacr}{\lambda_T}
= 10^{-6}L_{\rm edd}
\label{CRedd}
\ee  
Ultimately, the ratio $\lambdacr/\lambda_T$ may be thought of
as the inverse of the ratio of the effective cross section for a CR to
interact with a unit of plasma mass divided by the corresponding value
for a photon.

\subsection{Plan of this work}

In the next section we present the main results of this work.  First,
a brief overview of the observed properties of bright starbursting
galaxies is given in \S\ref{ss: observed}.  The basics of a
momentum-driven CR feedback mechanism is given in \S\ref{ss: maxL}.
In \S\ref{ss:comparison}, and to some degree in \S\ref{ss:
winds} as well, we weigh the advantages and disadvantages of the various
galactic feedback mechanisms and then compare them to our CR feedback
mechanism.  In \S\ref{s:CRinput} we discuss some of the 
relevant microphysical properties of CRs that are relevant
for CR feedback and in \S\ref{sss: Faber-Jackson} we speculate 
as to whether or not CR feedback is responsible for the 
Faber-Jackson (1976) relation.  We conclude in 
\S\ref{s: conclusion}.

\section{Cosmic Ray Feedback in Starbursting Galaxies}

Galactic scale CR-driven winds have been studied from a theoretical
point of view for many years (see e.g.  Ipavich 1975; Breitschwerdt et
al. 1991).  However to our knowledge, the connection between star
formation and wind power as a function environment was not considered.

Recently, CR feedback in galaxy formation has been considered by
Jubelgas et al. (2006).  They found that CR feedback can quench star
formation for low mass galaxies with velocity dispersions $\sigma\leq$
80 km/s.  In what follows, we argue that CR feedback potentially
chokes off star formation even for galaxies that are much larger,
where the velocity dispersion $\sigma$ can reach values up to $\sim
300$ km/s.  We believe that our conclusion differs from theirs due to
a difference in the handling of CR transport, a point that is
discussed in \S\ref{ss: destruction}.

Before we delve into the physics of CR feedback and transport, some of
the observed properties of starbursting galaxies are summarized.


\subsection{Observed Properties of Starbursting Galaxies}
\label{ss: observed}

As previously mentioned, galaxies and even starbursting
galaxies are dim in comparison to their
quasar counterparts.  Bright young and gas rich
starbursters possess characteristic bolometric luminosities of
$L_{\rm bol}\sim 10^{45}-10^{46}\,{\rm erg\, s^{-1}}$ at the
massive end.  Starbursters are powered by UV photons
emanating from the surface of massive stars in which
the overwhelming majority ($\sim 90-99\%$) of the energy
release is reprocessed into the infrared,
a result of UV opacity onto dust grains.

Although starbursting galaxies exhibit rather complex structure and
characteristic properties may vary greatly with location, some
representative values of the densities, outflow velocities, and length
scales can be obtained from observations.  Typical $\rm H_2$ number
densities for extreme starburst regions vary between $10^2-10^4\;\rm
cm^{-3}$.  These densities are averages over regions which extend
$\sim 100-500\; \rm pc$ from the nucleus (e.g., Downes \& Solomon
1998).  High velocity winds are a generic feature of starbursts, both
locally and at high redshift (Veilleux et al., 2005).  Outflow speeds
in excess of the escape velocity are observed irrespective of the mass
of the host (see e.g., Fig. 7 of Martin 2005).  Mass outflow rates can
also be estimated, although less robustly.  Outflow rates are
typically inferred to be anywhere from $\sim$ a few to $\sim 100\%$ of
the star formation rate (Veilleux et al. 2005; Martin 2005).  These
estimates are sensitive to the length scale used to convert the
observed gas column to the mass outflow rate, which is difficult to
determine.

There is also copious evidence for high velocity winds in
the spectra of several $z>1$ galaxies.  This evidence comes from
blue-shifted absorption lines and P Cygni
Ly$\alpha$ emission line profiles in the rest frame UV spectra
(Veilleux et al. 2005).  Among these high redshift galaxies,
the lensed source, MS 1512-cB58, has the highest quality spectra.
The bulk outflow velocity is $255 \; \rm km\; s^{-1}$, with
absorption extending to $\sim 750 \; \rm km\; s^{-1}$, well above
the escape velocity (Pettini et al. 2002).  These outflow velocities
are not uncommon, and are, in fact, typical of luminous Lyman
break galaxies (Adelberger et al. 2003, Shapley et al. 2003).



%
 
\subsection{Maximum Luminosity of Starbursting Galaxies}
\label{ss: maxL}

As previously mentioned, CRs are an energetically 
subdominant by-product of star formation.  But, due 
to their small mean free path with the magnetized 
interstellar medium, the collective momentum 
response with a given galaxy's gaseous component 
might have dramatic consequences.  In order to 
determine under what conditions CR production 
will heavily influence the dynamics of a galaxy, 
we consider galactic hydrostatic balance were
the CR force is balanced by gravity alone.  

For the sake of simplicity and a lack of theoretical 
and observational constraints, we take a one-zone 
approach.  That is, we collapse all of the information
regarding the structure of a given galaxy into
a single characteristic value taken at a characteristic 
length scale.  Later on, in \S\ref{ss:comparison} we
justify this approach.  Furthermore, from here on, we 
assume that the mass distribution of a given 
galaxy resembles that of a singular isothermal sphere whose 
relevant properties are briefly listed in Appendix 
\ref{a:isothermal}.   

In the interest of clarity, we derive the maximum luminosity in CRs --
and incidentally photons as well -- of young star forming galaxies by
examining hydrostatic balance in three different ways.

\subsubsection{Hydrostatic Balance}\label{ss:balance}

In order for the CRs to limit the inward 
flow of gas, thus quenching star formation, they 
are most likely responsible
for the bulk of the galactic pressure.  In the diffusion limit, 
the CR pressure is related to the CR flux by the relation
\be
\Pcr\simeq \Fcr\frac{\taucr}{c},
\ee
where $\taucr$ is the CR proton optical depth.
The CR energy flux $\Fcr$ is related to the star formation 
rate per unit area $\dot{\mu}_{\rm _{SF}}$ by the relation 
\be
\Fcr\simeq \epsilon_{_{\rm CR}}\dot{\mu}_{_{\rm SF}}c^2.
\ee
In the above expression $\epsilon_{_{\rm CR}}$ and $\dot{\mu}_{_{\rm SF}}$
are the efficiency of converting mass into CR proton energy and 
the starformation rate per unit area, respectively.
If we assume the galaxy in question resembles an isothermal 
sphere, the flux of CRs can be related to a star formation 
rate within a given radius
\be
\Fcr\simeq 6\times 10^{-3}\epsilon_{6}\dot{m}_{\rm _{SF}} 
R^{-2}_{\rm kpc}\,{\rm erg\,cm^{-2}\,s^{-1}}
\ee
where $\epsilon_{6}$, $\dot{m}_{\rm _{SF}}$, $R_{\rm kpc}$ is the efficiency of
converting rest mass into CR energy in units of $10^{-6}$, star formation
rate in units of solar masses per year, and galactic radius in units of 
kpc, respectively.  

Now, we may write the CR pressure as 
\be
\Pcr\simeq 3\times 10^{-13}\taucr\epsilon_6\dot{m}_{\rm _{SF}}
R^{-2}_{\rm kpc}
\,{\rm erg\,cm^{-3}}.
\ee
In our feedback picture, the outward 
diffusion of CRs halts gas accretion and
thus star formation once CR pressure is almost entirely supporting the
gaseous component of the galaxy against the inward force of gravity.  
In that case, hydrostatic balance dictates 
\be
\Pcr\simeq\rho\,\sigma^2\simeq 3\times 10^{-7}f_gR^{-2}_{\rm kpc}
\sigma^4_{300}
\ee
for an isothermal sphere.  With this, a criteria relating
star formation rate and CR optical depth to the mass of the galaxy, or 
equivalently, its velocity dispersion is given by
\be
\dot{m}_{\rm _{SF}}\simeq 10^{3}\,\tau^{-1}_3\epsilon^{-1}_6
f_{g_{0.1}}\sigma^4_{300},
\label{mdotsigma}
\ee
which does not explicitly depend upon galactic radius.  Note 
that $\tau_3$ and $f_{g_{0.1}}$ are the CR optical 
depth in units of $10^3$ and the gas fraction normalized to 
$0.1$, respectively.  The value for $\dot{m}_{\rm _{SF}}$ in the 
above expression seems enormous.  Interestingly, this is truly 
an upper limit.  

First, it is unlikely that CR feedback can affect a galaxy whose
velocity dispersion is much larger than $\sigma_{300}$ since only
$\sim 1{\rm keV}$ per baryon (which corresponds to a velocity
dispersion $\sigma\sim 300\,{\rm km s^{-1}}$) is available in the form
of CRs for an efficiency $\epsilon_6$.  From the arguments in Appendix
\ref{a:wind}, it follows that the maximum value of the optical depth
for a wind with a terminal velocity $v_w\sim \sigma_{300}$ is of order
$\taucr\sim 10^3$.  Furthermore, a galaxy with a velocity
dispersion $\sigma\simeq 300\,{\rm km s^{-1}}$ certainly represents
the high end of galactic mass and gravitational potential and thus,
one would not expect values of the star formation rate in excess of
eq. (\ref{mdotsigma}).

The result given by eq. (\ref{mdotsigma}) yields a maximum star
formation rate that is nearly identical to the value obtained by
eq. (19) of MQT as long as 
$\taucr$ and $\epsilon_{_{\rm CR}}$ are independent of galactic
mass and velocity dispersion.  In what follows, we explore the 
meaning of this apparent coincidence.


\subsubsection{(Cosmic) Radiation Pressure-Driven Wind}
\label{ss:wind}

Instead of balancing CR pressure with gravity, we now phrase 
hydrostatic balance in
terms of the CR luminosity and enclosed mass.  Again, by assuming spherical 
symmetry, hydrostatic balance is given by 
\be
\frac{\kappa_{\rm CR}}{c}\Fcr=\frac{\kappa_{\rm CR}}{c}
\frac{\Lcr}{4\pi r^2}\simeq \frac{GM(r)}{r^2}.
\ee
If we integrate over mass i.e., $dm=\rho\,d^3x=4\pi\rho r^2dr$,
then opacity may be replaced with optical depth.  We have
(see e.g., Lamers \& Cassinelli 1999)
\be
\Lcr\frac{\taucr}{c}\simeq \frac{f_g\,G\,M^2(R)}{R^2}
\label{Lsigma1}
\ee
If the mass distribution of the galaxy in question resembles an isothermal 
sphere, then we may write
\be
\Lcr\taucr\simeq \frac{4c\,f_g}{G}\sigma^4,
\label{Lsigma2}
\ee 
which is strikingly similar to the Faber-Jackson (1976)
relation.
In terms of typical scalings for galaxies
\be
\Lcr\taucr \simeq 3\times 10^{46}\,f_{g_{0.1}}
\sigma^4_{200}\,{\rm erg\,s^{-1}}
\label{Lsigma3}
\ee
where $f_{g_{0.1}}$ and $\sigma_{200}$ are the gas fraction and
velocity dispersion normalized to value of $0.1$ and $200\,{\rm km\,
s^{-1}}$, respectively.  Note that eq. (\ref{Lsigma3}) is identical to 
eq. (\ref{mdotsigma}).

For every gram of matter that is assembled into stars, CRs are
responsible for only $\sim 10^{-3}$ of the energy release while
photons are responsible for the overwhelming majority.  Consequently,
the CR luminosity is smaller than the starburst photon luminosity by a
factor of $\sim 10^3$.  Motivated by the assumption that every UV
photon released from the surface of a massive star is absorbed by dust
grains and subsequently downgraded into non-interacting infrared
radiation, MQT employ the ``single scattering approximation,'' which
sets the photon optical depth $\tau_{\gamma}=1$ for all galaxies.  If 
$\taucr\sim 10^3$ as in the case of the Milky Way, then the 
maximum star formation rate, CR luminosity, and photon luminosity 
given by eq. (\ref{Lsigma3}) is equal to the upper limit 
derived by MQT in the case where photon feedback determines the 
maximum galactic luminosity.  


Note, that we have yet to consider the dependencies of bulk
galactic properties such as the velocity dispersion $\sigma$ and 
gas fraction $f_g$ upon the 
optical depth $\taucr$ and efficiency $\epsilon_{_{\rm CR}}$.

\subsubsection{The Eddington Limit in Cosmic Rays}
\label{ss:edd}

Now, we exchange CR optical depth $\taucr$ with CR opacity
$\kappa_{\rm CR}$ in order to obtain an equivalent expression for the
maximum CR luminosity $\Lcr$.  In order to do so, we must
determine the effective cross section or mean free path that a typical
CR has with magnetized interstellar matter.

Let us take the 1-10 GeV CR mean free path to be
$\lambdacr\sim 1\,{\rm pc}$, approximately the Milky 
Way value, as a benchmark.  In order to obtain an Eddington 
limit, we compare the CR mean free path $\lambdacr$ 
to the Thomson mean free path $\lambda_{T}$ i.e., 
\be
L_{\rm edd, CR}\simeq L_{\rm edd}\frac{\lambdacr}{\lambda_T}
=4\pi GcM(r)\rho(r)\lambdacr
\label{CRedd2} 
\ee
or
\be
L_{\rm edd, CR}\simeq 3\times 10^{43}\lambda_{_1}
f_{g_{0.1}}\sigma^4_{200}R^{-1}_{\rm kpc}\,{\rm erg\,s^{-1}}
\ee
where $\lambda_{_1}$ is the CR mean free path in units of
a parsec.  

The arguments surrounding eq. (\ref{CRedd2}) are identical to the
Eddington argument put forth by Scoville (2003, hereafter SCO)
regarding UV photon opacity on dust.  His primary motivation was to
ascertain the supposed common origin of the light to mass ratio
$L/M\sim 500 L_{\odot}/M_{\odot}$ of both self-gravitating giant
molecular clouds in M51 and the central region of the dust-rich luminous
starburst galaxy Arp 220.

We define a CR to mass ratio 
${\mathcal R}_{\rm CR}\equiv \Lcr/M(r)$, which for an 
isothermal sphere is given by
\be
{\mathcal R}_{\rm CR}\simeq 2\,f_{g}\sigma^2_{200}R^{-2}_{\rm kpc}
\lambda_{_{1}} {\mathcal R}_{\odot},
\label{CR2mass}
\ee
where ${\mathcal R}_{\odot}=L_{\odot}/M_{\odot}\sim 2\,{\rm erg\,s^{-1}
g^{-1}}$ is the light to mass ratio of the Sun. If the efficiency 
of generating photon power from the act of star formation is
$\epsilon_{\rm _{SF}}/\epsilon_{_{\rm _{CR}}}\sim 10^3$, then the 
upper limit for the light to mass ratio of starbursting environment,
${\mathcal R}_{\rm SF}$, is given by
\be
{\mathcal R}_{\rm SF}\simeq \frac{\epsilon_{_{\rm SF}}}{
\epsilon_{_{\rm _{CR}}}}{\mathcal R}_{\rm CR}\simeq 
2\times 10^3\,f_{g}\sigma^2_{200}R^{-2}_{\rm kpc}
\lambda_{_{1}} {\mathcal R}_{\odot}.
\label{light2mass}
\ee

The gas fraction is roughly $f_g\sim 1/2$ for both sets of
environments considered by SCO.  For a CR mean free path
$\lambdacr\sim 1$pc, the above value for ${\mathcal R}_{ \rm SF}$ is
reasonably close to the measured value for Arp 220, given that the
velocity dispersion of Arp 220 is slightly different than
$\sigma_{200}$, the mass distribution is not perfectly described as an
isothermal sphere, and the observed $L_{\rm SF}$ (or the bolometric
luminosity) is only accurate to a factor of 2 or so.  Giant molecular
clouds have typical dimensions of $\sim 10-30$ pc and velocity
dispersion $\sigma \lesssim 10\,{\rm km\,s^{-1}}$.  Therefore,
according to the form of ${\mathcal R}_{\rm SF}$ given by
eq. (\ref{light2mass}), both systems will be limited to a similar
light to mass ratio resulting from CR feedback.  In Appendix
\ref{a:turb} we discuss the viability of {\it intrinsic} CR feedback
in giant molecular clouds.

\subsection{Cosmic Rays vs. Photons and Supernovae}\label{ss:comparison}

Now, we assess how the three major by-products of star formation
i.e., starlight, supernovae, and cosmic rays compare with one
another in terms of their ability to inject momentum into the 
interstellar medium.  Special attention is given when comparing
CRs and stellar photons.  Our arguments are given below.

\subsubsection{Global Kinematics of Various Feedback Mechanisms}

Energetically, cosmic rays are a highly subdominant by-product of star
formation.  However, if the injection of momentum determines whether
or not the star forming constituents irreversibly disrupt the galaxy
of their birth, then rate of energy injection may not, in fact, 
be determining factor.
At the crudest possible level of discourse, the rate at which 
momentum is deposited in a galaxy $\dot{P}$ is given by

\be
{\dot P}\sim {\dot{E}}/{v}\sim {L}/{v}
\ee 
where $v$ is the characteristic velocity at which  
the energy carrying particles are transported through the
interstellar medium.  

As previously mentioned, $L_{_{\rm SF}} > L_{_{\rm SN}} > 
\Lcr$.  However at the same time, $c>v_{\rm SN}>
v_{\rm CR}$.  In the case of core-collapse SNe, 
\be
\dot{P}_{_{\rm SN}}\sim {L_{_{\rm SN}}}/{v_{_{\rm SN}}}\sim 10^{33}
{\dot m}_{_{\rm SF}}\,{\rm g\,cm\,s^{-2}}
\ee     
where $v_{\rm SN}\sim 3\times 10^8\,{\rm cm\,s^{-1}}$ is the 
characteristic velocity of a supernova shock.

For cosmic rays, 
\be
\dot{P}_{_{\rm CR}} & \sim & \Lcr/v_{_{\rm CR}}\sim \Lcr\,
\taucr/c\\
& \sim & 3\times 10^{33}
\epsilon_{_{\rm CR}}\,\dot{m}_{\rm SF}\,{\rm g\,cm\,s^{-2}}
\nonumber ,
\ee  
which is slightly larger than the value 
obtained from supernovae.  Here, 
$v_{\rm CR}\sim H/t_{\rm diff, CR}\sim 10^7\,{\rm cm\,s^{-1}}$
is the characteristic velocity at which cosmic rays {\it diffuse}
in bulk.  Note that $v_{\rm CR}\sim c/\taucr$, where 
$\taucr$ is the optical depth in cosmic rays given 
by $\taucr\sim H/\lambdacr\sim 3\times 10^3$ for
the Milky Way.

In the case of starlight, 
\be
\dot{P}_{_{\rm SF}}\sim L_{_{\rm SF}}/c\sim 5\times 10^{32}\,
\dot{m}_{_{\rm SF}}\,{\rm g\,cm\,s^{-2}}.
\ee    
Therefore all three forms of momentum deposition are comparable.  
If the photons are optically thick, then 
$c\rightarrow c/\tau_{\gamma}$ in the above
expression where $\tau_{\gamma}$ is the photon optical depth.

\subsubsection{Cosmic Rays vs. Photons}
\label{sss: CRvPh}

Eqs. (\ref{mdotsigma}), (\ref{Lsigma3}) and (\ref{light2mass}) all
arrive at the same conclusion i.e., for a CR mean path $\lambdacr
\sim 1$pc and radiative efficiency $\epsilon_{_{\rm CR}} \sim 10^{-6}$
-- values that are reasonably close to the Milky Way value -- then,
the injections and transfer of 
1-10 GeV CRs will inevitably unbind a given galaxy's gaseous
component.  However, eqs.  (\ref{mdotsigma}), (\ref{Lsigma3}) and
(\ref{light2mass}) also inform us that for the same choice of
parameters, the basic observational signature of CR feedback -- a
maximum value for the equivalent physical quantities $\dot{m}_{\rm
_{SF}}$, $L_{\rm SF}$ and ${\mathcal R}_{\rm SF}$ -- simultaneously
indicate that it is photon-feedback, according to the theory of SCO
and MQT, that caps the luminosity of a given galaxy.

To our knowledge, there is no simple straightforward observational
test that distinguishes between the CR mechanism presented in this
work and the photon-driven theory.  However, a clear distinction can be
drawn on theoretical grounds alone.

The advantage of utilizing starburst photons as a feedback mechanism,
which limits star formation, is that the overwhelming majority of the
energy and power ($\sim 99\%$) from the action of assembling gas into
stars is released in the form of photons.  At the same time only $\sim
1/1000$ of the energy release and power from star formation comes out
in the form of CRs.  Furthermore, the coupling between starburst
photons and dusty matter is relatively well understood: UV starlight
is absorbed by dust grains and immediately downgraded into
non-interacting infrared radiation.  The frequency-weighted opacity
for this process is roughly a few hundred times the Thomson opacity
(SCO; MQT).  Finally, measurements of the bolometric photon luminosity
of gas-rich starbursting environments directly constrain any
photon-feedback theory, whereas for CR feedback a CR proton conversion
efficiency $\epsilon_{_{\rm CR}}$ must be assumed from the
starformation rate.

In order to unbind interstellar
matter in the largest and most massive galaxies, the CR mean free path
$\lambdacr\sim 1$pc or smaller, which implies that the CR
optical depth $\taucr$ must be large such that $\taucr
\sim 10^3$, or even higher.  Therefore, any theory of galaxy feedback
regulated by CRs requires, as in the case of the Milky Way, a large CR
optical depth.

We now raise a crucial point: { \it Due to their large
optical depth, each individual CR injects momentum into the
interstellar medium in the direction opposite a given galaxy's
gravitational center.}  Thus, the CR flux and pressure gradient do not
strongly depend on the spatial distribution of injection sites.
Furthermore, the fact that $\taucr$ must be large implies that
the one-zone model formulation invoked in
\S\S\ref{ss:balance}-\ref{ss:edd} is well justified at the order unity
level.
  
The momentum recoil of the interstellar medium resulting from the
absorption of UV starlight onto dust grains strongly depends on the
spatial distribution of the photon injection sites.  To
demonstrate this, we relax an important simplifying assumption employed
by SCO and MQT i.e., approximating that the entirety of a
given galaxy's UV radiative flux emanates from a single centrally
located point, behind a common dust sublimation layer.     
 
In the Milky Way, for example, there are $\sim 10^4$ O stars (Wood \&
Churchwell 1989) and for young stellar populations with a Salpeter
IMF, O stars dominate the radiative power.  For the sake of our
argument, we assume that each individual O star is contained within a
molecular cloud-like environment for the overwhelming majority of its
life.  Note that this is not the case in the Milky Way, but is likely
to be a good approximation in a gas-rich starbursting galaxy.  As
previously mentioned, UV starlight from embedded massive stars is
reprocessed and downgraded into the infrared by absorption onto dust
grains (Churchwell 2002).  It is this source of opacity upon which SCO
and MQT build their theory.  By doing so, they conclude that starburst
photons enjoy a relatively large momentum coupling with the gaseous 
component of their parent galaxy.

In Appendix \ref{a:breakdown}, we briefly outline why the one-zone
approximation -- meaning that the UV photons originate from a single
point -- of SCO and MQT qualitatively and quantitatively breaks down
on galactic scales.  The observational consequence is that starlight
can only ``stir up'' the gas to characteristic velocities of order
$v\sim 1/10$ of the velocity dispersion $\sigma$ on length scales of
order $\Delta x \sim 10 {\rm pc}$, respectively.  It follows, that the
starburst luminosity would have to be $\sim$ 100 times larger than
the maximum starburst luminosity $L_M$ given by MQT if the onset of a
photon-driven wind is to limit the luminosity of a galaxy.  If this
were the case, the bolometric luminosities of galaxies would be $\sim
10-100$ times larger than their respective quasars, which clearly
violates current observations.

More recently, Thompson et al. (2005) postulate that, on the basis of
observations of bright local starbursters such as Arp 220, prompt
starburst photons self-regulate the star formation rate via the
infrared dust opacity.  If the dust is optically thick to far IR
emission, this model has many of the benefits of an optically thick CR
feedback model. Specifically, the net force imparted by each photon is
amplified due to multiple scatterings and the momentum imparted is
insensitive to the source distribution.  In order to be optically
thick to the far IR, large galactic gas surface densities
($\Sigma_{g}\geq 1$) and large dust temperatures ($T_d\simeq 100$ K)
over the bulk of the starburst region, as indicated by some
observations of Arp 220, are required (Downes \& Solomon 1998; Soifer
et al.  1998).  In this case, it is possible that the infrared photon
pressure may exceed the CR proton pressure near the midplane,
especially if CR destruction becomes important (see \S \ref{ss:
destruction}).  However, this interpretation that the nuclear activity
in Arp 220 is powered by star formation has recently been called into
question (e.g. Downes \& Eckart 2007).


\subsection{Aspects of Galactic Scale CR-Driven Winds}
\label{ss: winds}

As stated in \S\ref{ss: observed}, starbursters 
display outflows with velocities significantly in 
excess of the escape speed $v_{\rm esc}$ or the 
velocity dispersion $\sigma$.   Interestingly,
line-driven winds of massive stars and dusty 
radiation-driven winds of asymptotic giant branch 
stars also display fast winds with velocities in 
excess of $v_{\rm esc}$ due to multiple scatterings,
outside the main body of the respective object.    

Radiation (or cosmic radiation) pressure driven
winds occur when the luminosity of the source 
approaches and surpasses the frequency-averaged 
Eddington limit i.e., when the radiation force 
exceeds gravity.  If the radiatively supported source 
is in quasi-hydrostatic balance beneath its photosphere, then 
a wind can only develop if the opacity increases above
the main body's photosphere.  For example, in the winds 
of AGB stars, the fact that dust sublimates at temperatures
below the photospheric temperature leads to an opacity 
increase with decreasing density (Salpeter 1974; Goldreich 
\& Scoville 1976; Ivezic \& Elitzur 1995).  

To see how this works, we write
\be
\rho\,\frac{\kappa\, F}{c}=\frac{F}{\lambda\,c} =\rho\,g
+\rho\,v\frac{dv}{dr}
\label{e: wind1}
\ee  
where $\kappa$, $\lambda$, $F$, $v$, and $g$ is the radiation opacity,
mean free path, flux, outflow velocity, and gravitational
acceleration, respectively.  If $\kappa$ increases with decreasing
density, then the ratio of radiative to gravitational acceleration
increases as a parcel of gas moves outwards.  Integrating over volume
allows us to write 
\be 
4\pi\int^{\infty}_{R_0}\,r^2dr\left(\frac{F}{\lambda\,c} -\rho\,g\right) 
\simeq\frac{L}{c}\tau_w=\dot{M}v_{\infty}
\ee
i.e., infinite acceleration may occur in the limit of infinite optical
depth.  Here, $\tau_w$ is the optical depth of the wind, $\dot{M}=4\pi
R^2\rho\,v$ is mass loss rate, $v_{\infty}$ is the terminal velocity,
and $R_0$ is the launching radius.  In the above expression, the
gravitational acceleration is neglected in the second term since it is
assumed to be small in comparison to the radiative acceleration.  Of
course, $\tau_w$ is limited by energy conservation, which we discuss in
Appendix \ref{a:wind}.

The scenario outlined above for radiatively-driven winds of stars may
analogously occur for CR-driven galactic winds.  Cosmic ray transport
in the Milky Way is often modeled with a constant diffusivity over a
large rarefied kpc halo; a picture that is consistent with
observations of the galactic CR distribution .  Since the diffusivity
varies little -- or not at all -- with density, the CR cross section
and thus opacity, increases with decreasing density.

In \S\ref{s:CRinput} we discuss how the CR mean free path and cross
section directly depend only upon the magnetic field fluctuations at
the Larmor scale.  That is, the level of coupling between CRs and
matter is, in principle, independent of column density -- an ideal
situation for winds with multiple scatterings.   For example, if the
streaming instability is the source of resonant magnetic fluctuations
(see \S\ref{sss: streaming}), then $\kappa_{\rm CR}\propto
\rho^{-1/2}/B$, where $\kappa_{\rm CR}$ is the CR ``opacity.''

As stated in \S\ref{sss: CRvPh}, UV photons released from the surface
of massive stars are absorbed and then quickly downgraded into
effectively non-interacting IR photons.  Therefore, $\tau_{\gamma}\sim
1$ for photon feedback -- if one accepts an idealized one-zone
approximation (see \S\ref{sss: CRvPh}).  When the momentum of the
radiation field is so weakly coupled to the matter, it is difficult to
imagine how photon-driven feedback leads to wind velocities in excess
of $v_{\rm esc}$ by a factor of $\sim $ a few, unless the light to
mass ratio exceeds that of eq. (\ref{light2mass}) by a factor of
$\sim$ 10 or so, since the maximum luminosity $L_M$ 
$\propto v^2_{\infty}$ for a fixed enclosed mass $M$
according to formalism of MQT (see e.g., their eq. (26)).

Independent of galactic mass, SNe shocks heat interstellar gas to
temperatures of about $\sim 1$keV, which corresponds to a sound speed
of $\sim 300$ km/s (Martin 1999, 2004).  If SNe lead to a
thermally-driven wind that is responsible for ejecting the interstellar
medium, then the outflow velocity is comparable the sound speed.  It
follows that thermally-driven winds from SNe are unlikely candidate
for explaining fast $\gtrsim 300$ km/s galactic winds.

\subsubsection{Upper Limit on $\taucr$ from Momentum and Energy 
Conservation }
\label{sss: wind_tau}

Even in our Galaxy, the interpretation of direct and indirect
measurements of CR transport may lead to qualitatively different
pictures as to how CR diffuse through the Milky Way and its
corona-like halo.  Nevertheless, we may set some rough limits on the
CR optical depth $\taucr$ without knowing the details of
the particular mechanism responsible for the resonant scatterings.

In the Appendix \ref{a:wind}, we briefly estimate an upper limit on
$\taucr$, following a well established result from the stellar
winds literature (Lamers \& Cassinelli 1999; Ivezic \& Elitzur 1995;
Salpeter 1974).  For optically thick radiation-driven winds, momentum
and energy conservation set an upper limit for the wind optical depth
roughly as $\taucr\sim c/v_w$, where $v_w$ is the wind terminal
velocity.

The maximum value of $\taucr$ takes on interesting values when
$v_w$ is large enough as to merit escape from a given galaxy's
gravitational potential.  As previously mentioned, since CRs inject
roughly $\sim 1 {\rm keV}$/baryon into the interstellar medium, the
deepest gravitational potential from which they can unbind gas
corresponds to a velocity dispersion $\sigma\sim 300\,{\rm
km\,s^{-1}}$.  If $v_w\sim \sigma$, then the corresponding optical
depth is $\taucr\sim 10^3$, which is close to the Milky value.

\section{Physical Properties of Interstellar Cosmic Rays}
\label{s:CRinput}

For lack of a better choice, we use the Galaxy in order to calibrate
the properties of cosmic ray injection, transport, and star formation
in general.  

Cosmic ray protons within the relatively modest energy scale of $\sim
1-10$ GeV are responsible for $\sim 90\%$ of the Galactic CR pressure
near the solar neighborhood.  Above 10 GeV, The CR energy energy
spectrum $n(E)$ is well described by a single power law such that
$n(E)\propto E^{-n}$ from $\sim 10-10^6$ GeV.  As already implied, the
CR spectrum is steep such that $n>2$, with a commonly quoted value of
$n\simeq 2.7$.

The spatial distribution of Galactic CRs is highly isotropic such that the
energy-weighted CR mean free path $\lambdacr\sim 0.1-1$ pc.
This remarkably low value for $\lambdacr$ is most likely 
the result of CRs scattering off of interstellar 
magnetic irregularities.  Then, for
typical Galactic scale height $H\sim 1$ kpc, the interstellar matter
of the Galaxy may be considered as possessing an optical depth to CRs
$\taucr\sim 10^3-10^4$.  Therefore, the CR pressure $\Pcr$ 
is related to the CR flux $\Fcr$ by the simple relation
\be
\Pcr\simeq \Fcr\frac{\taucr}{c}
\ee
over a CR pressure scale height.

In the Galaxy, a core-collapse
supernova occurs once every century, leading to a core-collapse supernova
luminosity $L_{\rm SN}$
\be
L_{\rm SN}\sim 10^{51}{\rm erg}/10^2{\rm yrs}\sim 3\times 10^{41}
{\dot m}_{\rm _{SF}}\,{\rm erg\,s^{-1}}
\ee
where ${\dot m}_{\rm _{SF}}$ is the normalized star formation 
rate in units of $1 M_{\odot}/{\rm yr}$.  The injection of CRs 
results primarily from 
the interaction of supernova shocks with the interstellar medium 
(Ginzburg \& Syrovatskii 1964).  However, CR deposition from other 
types of starburst outflows, such as stellar winds from massive 
stars, may also 
significantly contribute (Schlickeiser 2002).  In 
terms of the star formation rate $\dot{m}_{\rm _{SF}}$ and 
an CR efficiency $\epsilon_{_{\rm CR}}$, we may write the 
CR luminosity as 
\be
\Lcr=\epsilon_{_{\rm CR}}\dot{m}_{\rm _{SF}}c^2=
6\times 10^{40}\,\epsilon_{6}\,{\dot m}_{\rm _{SF}}
\,{\rm erg\,s^{-1}}.
\ee  
where $\epsilon_6$ is equal to $\epsilon_{_{\rm CR}}$ in units
of $10^{-6}$ or in other words, $\Lcr$ is normalized to 
$\simeq 20\%$ of $L_{\rm SN}$.

\subsection{Observational Constraints on $\lambdacr$
from the Leaky Box and Galactic Halo Models}
\label{ss: CR_obs}

Upper limits on the anisotropy of the CR distribution function, the
relative abundances of primary to secondary CRs, and the abundance
ratios between the radioactive isotopes of a given CR species sum up
the directly available observations of CR proton transport (Schlickeiser
2002).  Indirect tracers of CR behavior include radio emission,
presumably powered by bremsstrahlung and synchrotron losses of CR
electrons, as well as gamma-ray emission resulting from the decay of
mesons produced in proton-proton collisions.  Altogether, these
observational constraints indicate that CRs stay in contact with the
Milky way for time scales in the neighborhood of $10^7$ yrs.

Though the direct and indirect observations of CR transport dictate
that CRs scatter $\taucr^2\sim 10^6-10^8$ times before leaving the
Milky Way, they do not uniquely determine the volume of the region in
which they are confined (see e.g., Ginzburg et al. 1980 for a
discussion).  In the one-zone model, or ``grammage'' formulation, two
important pieces of information can be extracted; the galactic
residence time for a CR proton of a given energy and the average
hydrogen number density $\left<n_{\rm H}\right>$ during its encounters
with the interstellar medium.  For $\sim 1-10$ GeV CR protons $\left<
n_{\rm H}\right>\sim 0.2\,{\rm cm^{-3}}$ -- significantly smaller than
average value for the $\sim 100\,{\rm pc}$ scale height gaseous disk,
where is $n_{\rm H}\sim 1{\rm cm^{-3}}$.  In the grammage formulation,
typical residence timescales in the disk are of order $\sim 3\times
10^6\,{\rm yrs}$.  Together, this implies an optical depth in the disk
of $\taucr \sim 10^4$ and a mean free path of order $\lambdacr\sim
0.1{\rm pc}$.

The average hydrogen number density $n_{\rm H}\sim 1\,{\rm cm^{-3}}$ 
for the 100 pc disk, which implies CR
protons must somehow avoid regions of relatively large density while
maintaining a small level of spatial anisotropy, of order one part in
$10^3-10^4$.  The mass distribution of the Milky Way is partitioned 
into several phases of various densities and temperatures.  
Even if CR protons mainly restrict themselves to a phase with 
characteristic hydrogen number density $n_H\sim 0.2\,{\rm cm^{-3}}$,
it seems unlikely that such a phase of the interstellar medium 
is homogeneously distributed in space at the one part in 
$10^3$ level, in the solar neighborhood.

In light of the difficulties with the one-zone interpretation of
grammage measurements, another scenario that is often used to
interpret grammage measurements is the disk + halo model (Ginzburg
\& Syrovatskii 1964; Ginzburg et al. 1980).  By modeling
the vertical profile of the Galaxy, the quantitative interpretation of
a CR proton's trajectory is much different in comparison to one-zone
disk models.  Here, 1-10 GeV CR protons are thought to be rather
uniformly distributed in a diffuse $\sim$ 1 kpc disk with typical hydrogen
number densities $n_{\rm H}\sim 0.1\,{\rm cm^{-3}}$, with a
characteristic escape time of order $\sim 10^7\,{\rm yrs}$ (Strong 
\& Moskalenko 1998; Webber \& Soutoul 1998) .  In this
picture the CR mean free path $\lambdacr\sim 1\, {\rm pc}$.

\subsection{Constraints on $\epsilon_{\rm CR}$ and 
$\lambda_{_{\rm CR}}$ from other Galaxies}

Although the mechanism responsible for the 1-10 GeV cosmic ray mean
free path in the Milky Way is not completely understood, it seems that the
value of the cosmic ray mean free path $\lambda_{\rm CR}$ may be
roughly the same from galaxy to galaxy, over a large range of
star formation rate.  Important constraints on cosmic ray propagation
come from studies of the tight far-infrared/radio correlation, which
spans over many decades in star formation rate (see e.g. Condon 1992).
The far-infrared (FIR) emission is due to optical/UV starburst photons
being reprocessed and down-graded into the FIR by dust, while the GHz
synchrotron emission is powered by 1-10 GeV cosmic ray electrons.
Like cosmic ray protons, cosmic ray electrons are supposedly
accelerated at the shock fronts of supernovae and stellar winds, by a
common Fermi acceleration mechanism.

Above an energy of 1 GeV, where both electrons and protons are 
relativistic, the CR mean free path of both species are the same
as long as the scattering opacity results from small scale 
resonant magnetic irregularities.  Therefore, a measurement of the 
1-10 GeV CR electron mean free path serves as a good proxy for the 
1-10 GeV CR proton mean free path.  By setting the value of the 
magnetic energy density approximately equal to the total cosmic ray 
energy density -- the so-called minimum energy argument -- the
total CR electron energy content within the synchrotron cooling radius 
is extracted.  Along with the total synchrotron power within 
the cooling radius, this yields the synchrotron cooling time for the 
1-10 GeV CR electrons.\footnote{We do not understand why the {\it total}
(electron+proton) cosmic ray energy density is utilized in the 
traditional minimum energy argument, since synchrotron emission 
is produced only by cosmic ray electrons.  Nevertheless, the assumption 
that the magnetic energy density scales with the cosmic ray proton 
energy density does not appreciably alter the value of the 
synchrotron cooling time as long as the the ratio of cosmic ray 
proton to electrons does not significantly exceed the Galactic
value.}  Together with the cooling time, the characteristic 
emission scale allows for a determination of the 
the 1-10 GeV CR mean free path, which is typically close to 1 pc
or in other words, close to  Milky Way value (Condon 1992).

As previously stated, the interaction of energetic outflows 
from massive stars (supernova and stellar winds) with their 
ambient gaseous environment is commonly thought to be the 
site of interstellar CR production.  In our galaxy, roughly
$10\%$ of the core-collapse supernova power is converted into
1-10 GeV CR proton power, $L_{_{\rm CR}}$, and of that only
$\sim 1\%$ of $L_{_{\rm CR}}$ manifests itself in 1-10 GeV 
CR electron power.  Under the assumption that all of 
the GHz radio emission from starforming galaxies results
from synchrotron cooling of CR electrons and that all of the FIR
emission is due to reprocessed starlight onto dust, the tight 
FIR-radio correlation indicates that the efficiency of 
producing CR electrons, per unit of starformation, is 
remarkably constant from galaxy to galaxy.  Finally, if a magnetized
Fermi shock acceleration process is responsible for producing
interstellar CRs from a bath of thermal particles at a common 
sub-relativistic energy, then one expects on theoretical 
grounds that the ratio of CR protons to electrons to be $\sim 100$      
or so, close to the observed value in the interstellar medium 
(Schlickheiser 2001).  Altogether, this implies that the 
efficiency of producing 1-10 GeV CR protons $\epsilon_{_{\rm CR}}$ 
varies little from galaxy to galaxy.

\subsection{The Source of Small Scale Resonant 
Magnetic Irregularities or: A Great Uncertainly in Cosmic Ray Physics}
\label{ss: uncertainty}

\subsubsection{Streaming Instability}
\label{sss: streaming}

CRs flow freely along large scale laminar magnetic fields as they
tightly gyrate about their guiding centers.  In the presence of small
scale magnetic fluctuations, the CRs themselves can amplify waves in
which they are resonant, a phenomena known as the streaming
instability (Kulsrud \& Pearce 1969; Wentzel 1974).  The phase
velocity of the waves in question (e.g. fast or Alfv\`enic waves) is
much smaller than the speed of light, the velocity at which the CRs
propagate.  Thus, a CR perceives a wave packet of magnetic disturbances as
stationary in time, but not space (Kulsrud 2005).  Therefore the
appropriate resonance condition for the interaction between a CR beam
centered in energy at some Lorentz factor $\gamma$ and a wave packet
of magnetic disturbances of some width $\Delta k$ is that the Larmor
radius of the CRs equal to the component of the wavelength projected
along the laminar field.

The growth time of the streaming instability is extremely short
in comparison to local dynamical times of interstellar matter for 1-10
GeV CRs.  In the absence of damping CRs excite waves which propagate 
along the background field with a growth rate proportional to the 
ratio of the streaming or diffusion velocity to the Alfv\`en speed
$v_D/v_A$.  As the amplitude of the waves grow, forward momentum is 
extracted from the streaming CRs as they are isotropized, while 
simultaneously reducing their streaming velocity.  Wave 
excitation continues until $v_D\sim v_A$, where the instability 
is quenched and the instability's source of free energy has been 
removed.
 
Now, we may write the CR optical depth as
\be
\taucr\simeq \frac{c}{v_A}.
\label{e:tau1}
\ee   
From this, the CR optical depth of the Milky Way halo with 
$v_A\sim 150\,{\rm km\,s^{-1}}$ is $\taucr\sim 2\times 10^3$,
which is roughly the observed value.  

 
In order for the streaming instability to operate at given resonant
wavelength, the growth rate of the instability must surpass the
individual damping rates resulting from all sources of dissipation,
such as ion-neutral drag.  The growth rate due to the streaming
instability for a given resonant Alfv\`en wave is proportional to the
number density of CRs $n(E)$ resonant with the magnetic disturbance 
in question.  Due to the steepness of the CR energy spectrum, 
the CR number density at 100 GeV $n(E=100\,{\rm GeV})$ is 
sufficiently low that driving resulting from the streaming of CRs
cannot overcome ion-neutral drag (Cesarky 1980).  The argument 
presented above represents a major shortcoming of the
``self-confinement'' picture of CR transport.

\subsubsection{Kolmogorov Cascade}
\label{sss: cascade}

On its own, the interstellar gas may be a rich source of small scale
resonant Alfv\`enic disturbances.  In this case, the CR mean free path
$\lambdacr$ is limited by the amplitude of magnetic
fluctuations on the Larmor scale.

In the frame of the CR beam, magnetic 
irregularities of the interstellar medium are effectively 
stationary and thus, diffusion occurs only in
space since energy is roughly conserved per scattering event.
Spatial diffusion proceeds in pitch-angle with 
disturbances that are resonant with the individual CRs in 
momentum-space.  The appropriate diffusion coefficient ${\mathcal 
D}_{\rm CR}$ is roughly given by
\be
{\mathcal D}_{\rm CR}\simeq c\,r_L\frac{B^2}{\delta B^2\left( 
r_L\right)}.
\ee 
Here, $\delta B\left( r_L\right)$ is the amplitude of magnetic
fluctuations on the Larmor scale.  If the magnetic fluctuations
follow a Kolmogorov scaling, then $\delta B =\delta B_0
\left(\lambda/\lambda_0\right)^{1/3}$, where $\lambda$, $\lambda_0$,
and $\delta B_0$ is the eddy scale, the stirring scale, and the 
magnetic fluctuation amplitude at the stirring scale.

The above expression reflects the notion that a given CR can 
only wander in pitch-angle of its guiding center with the aid of 
resonant magnetic irregularities.  Since the phase of the 
irregularities are taken as random, a CR must execute 
$B^2/\delta B^2(r_L)$ gyrations before its net trajectory,
and thus momentum, is altered at the order unity level.  It
follows that the CR mean free path $\lambdacr\simeq 
{\mathcal D}_{\rm CR}/c$, when the source of magnetic 
deformations is due to turbulence.  With this, the CR 
optical depth in the presence of resonant Kolmogorov magnetic turbulence 
reads
\be
\taucr  \simeq  H/\lambdacr\sim\,c\,H/{\mathcal D}_{\rm CR}
\,\,\,\,\,\,\, \nonumber\\
\taucr  \simeq \xi^2_B\left(\frac{H}{r_L}\right)^{1/3}
\left(\frac{H}{\lambda_0}\right)^{2/3}
\ee
where $\xi_B\equiv \delta B_0/B$.  

As previously mentioned, the Alfv\`enic cascade must possess, with
sufficient amplitude, eddies that are resonant with the CRs at the
Larmor scale.  In order to be resonant, the Alfv\`en waves should
propagate primarily along the field.  However, it is now generally
accepted that, in accordance with Goldreich, Sridhar and Lithwick's
(Goldreich \& Sridhar 1995; Lithwick \& Goldreich 2001) theory of
interstellar MHD turbulence, Alfv\`enic power flows only into eddies
that propagate in the direction perpendicular to the local large scale
field.  Thus, for $\lambda_0/r_L\gg 1$, there may be virtually no
resonant turbulent power left at the Larmor scale.
    
\subsection{Cosmic Ray Destruction}
\label{ss: destruction}

Cosmic rays can be destroyed once they interact with an ambient proton,
subsequently producing pions.  At 
a few GeV, the cross section for this inelastic process is $\sim 30
\,{\rm mb}$.  Therefore, the characteristic destruction time $t_{\pi}$ 
for GeV CR protons is (Schlickeiser 2002)
\be
t_{\pi}\simeq \frac{10^8}{n_{\rm H}}{\rm yrs},
\ee
which should be compared to the diffusion (or escape) time 
over a galactic scale height $H$ that is occupied with gas 
of average density $n_{\rm H}$.  

In the Milky Way, the destruction timescale $t_{\pi}$ does not 
play an appreciable role in the transport of CR protons.  Note that
this statement is independent on whether or not the one-zone or 
disk + halo model is utilized to interpret grammage measurements 
since the hydrogen number density throughout the Milky Way is 
relatively low.  

In young bright starbursting galaxies $n_{\rm H}$ may reach values of
$10^3-10^4\,{\rm cm^{-3}}$, corresponding to a destruction time of
$t_{\pi}\sim 10^5-10^4\,{\rm yrs}$.  It follows that in some of the
interesting cases in which CR feedback might be important, CRs may in
fact be destroyed before they traverse a galactic scale height.  In
this case, it is useful to employ an effective optical depth when
diagnosing the level of feedback that CRs exert back upon the
interstellar medium.  If a CR is destroyed before it escapes the
galaxy, it still has the the opportunity to exchange momentum with a
given galaxy's gaseous component.

In order to understand the effects of CR destruction, we have 
calculated the vertical variation of CR pressure for a homogeneous 
disk geometry in Appendix \ref{a: yukawa}. As in the case of scattering
dominated, radiative atmospheres (see e.g. Felten \& Rees 1972), the
physics is easily understood in terms of random walk arguments.
We define the effective CR optical depth, 
$\overline\tau_{_{\rm CR}}$, as
\be
\overline{\tau}_{_{\rm CR}}=\tau_{_{\rm CR}}
\sqrt{\frac{t_{\pi}}{t_{\rm diff}}}
=\sqrt{\frac{\tau_{_{\rm CR}}}{\tau_{\pi}}},
\label{e:tau_eff}
\ee 
where $\tau_{_{\rm CR}}\simeq H/\lambda_{_{\rm CR}}$, $t_{\rm diff}$ is
the diffusion time for the CR protons over the galactic scale height
$H$, and $\tau_{\pi}$ is CR the ``absorption'' optical depth due to 
pion production. 

The above expression acknowledges the fact that CRs must scatter
$\tau^2_{_{\rm CR}}$ times in order to traverse a distance $H$ in a
diffusion time $t_{\rm diff}$.  In the limit that $t_{\pi}< t_{\rm
diff}$, the square of the effective optical depth
$\overline{\tau}^2_{_{\rm CR}}$ is equal to the number of times a CR
proton scatters and exchanges momentum with the galaxy before its
destruction.

For sufficiently dense and extended systems, destruction of CR protons
due to pion production will inhibit the efficiency of CR feedback.
Consider the example of a gas-rich disk with $n_{H}\sim 10^4\,{\rm
cm^{-3}}$ and a disk scale height $H\sim 100\,{\rm pc}$.  The
destruction timescale $t_{\pi} \sim 10^4\,{\rm yrs}$ is quite short.
If $\tau_{_{\rm CR}}\sim 10^3$ for this $100$ pc disk then
$\lambda_{_{\rm CR}}\sim 0.1$ pc and $t_{\rm diff}\sim 3\times
10^5\,{\rm yrs}$.  Therefore, $\overline{\tau}_{_{\rm CR}}\sim 0.2
\,\tau_{_{\rm CR}}\sim 200$.
   
For galactic environments in which 1-10 GeV CRs are destroyed by
hadronic processes before their escape eq. (\ref{e:tau_eff}) provides
a prescription for estimating the CR contribution to hydrostatic
balance. If we assume the CR source is independent of height in the
disk, then pressure gradient (force) at the surface is given by
\be
\frac{d P_{_{\rm CR}}}{dz}=\frac{F_{_{\rm CR}} \overline{\tau}_{_{\rm CR}}}{H c}
=\frac{F_{_{\rm CR}}}{ \lambda_{_{\rm CR}} c}\sqrt{\frac{t_{\pi}}{t_{\rm diff}}}.
\label{e:dpdz}
\ee
Deeper within the atmosphere the force is smaller due to cancellation
of momentum between upward and downward moving CRs, going to zero (by
symmetry) at the midplane.

In situations were the CR mean free path $\lambda_{_{\rm CR}}$ is
small and the hydrogen number density $n_H$ is high, some 1-10 GeV CRs
may never escape the galaxy since meson production will destroy them.
The typical vertical displacement a CR drifts before its demise
${l_{_{\rm CR}}}$ is given by $l_{_{\rm CR}}\simeq \lambda_{_{\rm
CR}}N^{1/2}_s$, where $N_s=t_{\pi}/t_s$ is the number of scattering
before destruction and $t_s=\lambda_{_{\rm CR}}/c$ is the time in
between scattering events.  Rewritten in terms of $l_{_{\rm CR}}$, eq
(\ref{e:dpdz}) becomes
\be
\frac{d P_{_{\rm CR}}}{dz}=\frac{F_{_{\rm CR}}}{ \lambda_{_{\rm CR}} c}
\left(\frac{l_{_{\rm CR}}}{H}\right)
\label{e:dpdz2}
\ee
where we have implicitly assumed $l_{_{\rm CR}} < H$ (i.e. CRs are
typically destroyed before leaving the galaxy).  In the absence of CR
destruction processes, the term in parentheses would be unity.  The
effect of CR destruction has a simple interpretation.  Only those CRs
produced within $l_{_{\rm CR}}$ reach the surface before being
destroyed.  Therefore, only the fraction $l_{_{\rm CR}}/H$ of the CRs
contribute to the force at the surface.

A useful estimate of $l_{_{\rm CR}}/H$ can be obtained from the
relation $n_H \sim \Sigma_g/(m_p H)$ where $\Sigma_g$ is the gas
surface density. We then find that
\be
\frac{l_{_{\rm CR}}}{H} \sim \left(\frac{60 \rm \; g \; cm^{-2}}{\Sigma_g}
\frac{\lambda_{_{\rm CR}}}{H} \right)^{1/2}.
\ee
It is clear from this expression that if either $\Sigma_g$ is
sufficiently large and/or $\lambda_{\rm CR}/H$ is sufficiently small,
most CRs will be destroyed before escaping the starburst.  Of course,
reducing $\lambda_{\rm CR}$ at fixed $H$ produces a net {\it increase}
in the force due to the $\lambda_{\rm CR}^{-1}$ dependence in
eqs. (\ref{e:dpdz}) and (\ref{e:dpdz2}). However, this force is larger
only in an increasingly smaller fraction ($l_{_{\rm CR}}/H$) of the
volume near the surface.  Deeper in the atmosphere, a CR feedback
scenario will run into some of the same problems photon-driven
feedback suffers from as mentioned in Appendix \ref{a:breakdown}.
Near the surface there will be a force comparable to our estimate in
eq. (\ref{e:dpdz}), but throughout most of the starburst volume the
force due to upward moving CRs from any one source is mostly 
canceled by downward moving CRs from neighboring sources above.

Finally, we note that in their eq. (44), Jubelgas et al. (2006)
include the effects of CR feedback by assuming a pressure pressure
$P_{_{\rm CR}} \propto t_{\pi}$.  This gives a sharp reduction in
pressure with increasing density since $t_{\pi} \propto n_H^{-1}$. As
discussed in Appendix \ref{a:  yukawa}, this assumption is
appropriate deep within galaxy if $l_{_{\rm CR}}$ is much less than
the characteristic galactic radius.  However, in the outer regions
(within $\sim l_{_{\rm CR}}$) the CR forces will be significantly
larger due to the $d P_{_{\rm CR}}/dz \propto \sqrt{t_{\pi}}$
dependence in eq. (\ref{e:dpdz}).

\subsection{Overall Picture and Summary}

If in fact the expulsion of CRs is the mechanism responsible for
limiting the luminosity of starbursting galaxies, then the CRs must be
well-coupled to the interstellar gas such that the CR optical depth
$\tau_{_{\rm CR}}$ is large.  Of course, there are no direct
measurements of CR properties in other galaxies, which is why we have
primarily utilized observations and theories of the CR transport in
the Milky Way as our benchmark.  Even for the Milky Way, the value of
the CR mean free path $\lambdacr$ extracted from the data may vary by
an order of magnitude, where $\lambdacr\sim 1$pc if the halo model is
utilized and $\lambdacr\sim 0.1$pc for the leaky box interpretation.

From the theoretical side, the two most commonly accepted explanations
for the source of small scale resonant magnetic perturbations possess
severe weaknesses.  Since the growth rate of the streaming instability
at a given CR energy is proportional to $n(E)\propto E^{-2.7}$, the
streaming instability is not strong enough to overcome ion-neutral
damping at energies above 100 GeV (Cesarsky 1980).  Perhaps one might
account for the lack of an-isotropy -- and thus small mean free path
-- above 100 GeV asserting that the source or magnetic irregularities
on those scales comes about from a turbulent cascade of resonant
magnetic power.  However, it seems unlikely that the slope of the CR
distribution function with energy remains constant over $\sim 6$
decades in energy if the streaming instability were to inject power at
energies below $\sim 100$ GeV.

Despite these disturbing inconsistencies in the overall
picture of CR transport in the Milky Way, what we know 
for certain is that the scattering optical depth $\taucr$
is large such that $\taucr\sim 10^3-10^4$.  Furthermore,
we may conclude, with an acceptable degree of confidence, 
that the origin of the small CR mean free path $\lambdacr
\sim 0.1-1$pc in the Milky Way is magnetic in nature.  Within the 
context of galaxy formation, the fundamental issue is 
constraining how $\taucr$ and $\lambdacr$ 
evolves with environment (redshift and halo mass).  In what 
follows, we briefly discuss the manner in which $\taucr$
might evolve from galaxy to galaxy.

\subsubsection{Does Cosmic Ray Feedback Result in the 
Faber-Jackson Relation?  Maybe.} 
\label{sss: Faber-Jackson}

Eq. (\ref{Lsigma2}) appears strikingly similar to Faber-Jackson (1976)
relation.  Under the plausible assumption that galaxies residing on
the Faber-Jackson relation are faded and elderly versions of their
bright former starbursting past, then a momentum-driven feedback
mechanism may indeed lead to the Faber-Jackson relation.  If CR
feedback is responsible for the relation, then the CR optical depth
$\taucr$ can only evolve weakly with velocity dispersion $\sigma$.

The CR optical depth $\taucr$ depends on galactic magnetic
field strength if either the streaming instability or
a Kolmogorov cascade is responsible for the required resonant 
magnetic fluctuations.  The situation is more complicated in the case
where a magnetic cascade is responsible for the CR optical depth since
knowledge of $\xi_B$, $\lambda_0$, $H$ and the power law index of 
the cascade (= 1/3 for Kolmogorov) is required as well.  At the moment, 
a first principles calculation of $\taucr$ in the Milky Way 
would be a difficult task for reasons already mentioned.  Certainly,
an attempt to determine $\taucr$ from basic
physical arguments in other galaxies is out of reach. 

As stated in \S\ref{sss: wind_tau}, $\taucr$ cannot assume and
indefinitely high value.  In other words, only a finite number number
of scatterings are required before the CR beam vanishes due to
adiabatic losses and thus, energy conservation enforces an upper limit
on $\taucr$.  Furthermore, if the CR residence time, and
consequently $\taucr$, becomes too large then meson production
will limit the effective optical depth $\overline\taucr$ to a
moderate value.  Note that the $\taucr$ is bounded from above
as a result of constraints that have nothing to do with the ultimate
-- and highly uncertain -- mechanism that generates the required
resonant Alfv\`enic disturbances that scatter the CRs.

On less solid grounds, we can conclude that $\taucr$
cannot take on values that are too low for a given galaxy.  
If the galactic magnetic field is too weak, then the 
growth rate of the streaming instability is enormous 
since the ratio $c/v_A$ becomes very large.  On the other hand, if
the galactic field is quite strong on the stirring scale,
then one would expect that the amount of resonant power 
at $r_L$ is quite large, leading to a relatively small mean free 
path.  

\section{Conclusion}
\label{s: conclusion}

From the process of star formation, cosmic rays represent only 
a minor form of energy release.  Yet, their momentum coupling 
with the interstellar medium is absolute as result of their
small scattering length.  As long as the CR mean free path 
$\lambdacr$ does not greatly differ from the 
inferred value in the Milky Way, the collective force 
that CRs exert upon the interstellar medium as they leak out
is great enough to support the gaseous component against 
gravity and may even unbind it.  

If CR feedback is an important mechanism in determining the maximum
luminosity of a starbursting galaxy, then the CR optical $\taucr$ 
must be large in order to compensate for the relatively small CR
luminosity.  Consequently, galactic CRs necessarily impart momentum upon
the interstellar medium in the direction opposite to the galaxy's
gravitational center.  Thus, CR feedback is
highly independent of the spatial distribution of CR injection 
sites such as SNe and stellar winds.  

In analogy with the photon-driven winds of massive and 
asymptotic giant branch stars, CR-driven galactic winds 
may reach terminal velocities in excess of the escape velocity 
by a factor of $\sim$ a few.  Due to multiple scatterings
in the galactic halo, above the main body of the galaxy, 
the CR beam may indefinitely accelerate a parcel of fluid 
up to the point where all of the CRs are redshifted away.
   
The main uncertainty in the CR feedback theory presented here is
in determining the manner in which the CR mean free 
path $\lambdacr$ (or optical depth $\taucr$) 
and radiative efficiency $\epsilon_{_{\rm CR}}$ varies from 
galaxy to galaxy.  The optical depth $\taucr$
cannot approach arbitrarily high values since adiabatic 
losses and meson production will eventually remove 
energy out of the CR beam.

\acknowledgements{We thank P. Goldreich, M. Krumholz, R. Kulsrud,
J. Ostriker, E. Quataert, B. Paczy\'nski, C. Pfrommer, M. Rees,
A. Shapley, J. Stone and T. Thompson for helpful conversations.  AS
acknowledges support of a Hubble Fellowship, administered by the Space
Telescope Science Institute.  SWD and ERR acknowledge support of a
Chandra Fellowship administered by the Chandra X-Ray Center at the
Harvard-Smithsonian Center for Astrophysics.}
      
\appendix

\section{Appendix A: A galaxy as an isothermal sphere}
\label{a:isothermal}

As a baseline model, we assume the mass profile of a given 
galaxy is that of a singular isothermal sphere.  This assumption 
is more accurate for hot stellar systems such as elliptical galaxies 
and bulges and less accurate for disk galaxies.  The density 
profile of the gaseous component is given by 
\be
\rho\left(r\right)=\frac{f_g\sigma^2}{2\pi Gr^2},
\ee
which corresponds to an enclosed gas mass of
\be
M_g\left(r\right)=\frac{2f_g\sigma^2r}{G}.  
\ee
In the above expressions, $f_g$, $\sigma$, and $r$ are the 
gas fraction, velocity dispersion, and galactic 
radius, respectively.

\section{Appendix B: Breakdown of the one-zone model 
for radiative feedback}
\label{a:breakdown}

We assume that every massive star is embedded such that every single
one of its emitted UV photons are absorbed by dust grains and
reprocessed into infrared radiation close to the star.  This is not
the case in the Milky Way where massive stars are embedded for only
$\sim 10\%$ of their lifetimes (Wood \& Churchwell 1989).  However,
for starbursting sources, the ratio of infrared to UV radiation is
large, implying that our assertion of every massive star being
embedded is correct.

UV photon power is converted into the IR at a characteristic radius of
$\sim 0.1-1{\rm pc}$ from the center of each individual star.  It is
in this region, the so-called ``cocoon,'' in which momentum is
directly exchanged between light and matter.  Note that the
characteristic hydrogen number densities in this region range from
$n\sim 10^5-10^7\, {\rm cm^{-3}}$ (Koo \& Kim 2001; Churchwell 2002).

There are $\sim 10^4$ O stars in the Milky Way.  For a starbust galaxy
with $\dot{m}_{\rm _{SF}}\sim 100$ or so times the Milky way value,
one expects there to be $N_{O*}\sim 10^6$ or so massive stars
responsible for both the starburst luminosity as well as the injection
of radiation momentum.  If the volume of this galaxy is $\sim 10^9\,
{\rm pc^3}$, then the typical separation between each O star is
$\Delta x\sim \left(V/N_{O*}\right)^{1/3}\sim 10\,{\rm pc}$ for a
homogeneous and spherically symmetric distribution of O stars.

Furthermore, the photons from each O star may be thought of as
being responsible for injecting momentum into $1/N_{O*}$ of the given galaxy's 
gaseous component.  It follows, that momentum balance for a parcel of dusty 
gas placed somewhere between two typical massive stars may be expressed
as 
\be
\rho\frac{\kappa_{\rm UV}}{c}\frac{L_{O*}}{4\pi\Delta x^2}\sim \rho
\frac{v^2}{\Delta x/2},
\label{e:stirring}
\ee
where $\kappa_{\rm UV}$ and $L_{O*}$ is UV opacity on dust and the 
characteristic luminosity of an O star.  In the above expression, the 
radiation force imparted at the surface of the dusty ``cocoon'' is
matched by a non-linear momentum sink, which can be thought of as either
an outward moving shell or a parsec-scale turbulent eddy.  Integrating over
the volume in between the two stars gives
\be
\frac{L_O*}{c}\,\tau\sim M_{\Delta}\frac{v^2}{\Delta x/2},
\ee
where $M_{\Delta}\sim f_g\,M(R)/N_{O*}$ is the fraction of gaseous 
galactic mass in which a given massive star is responsible for 
accelerating.  Noting that $L_{O*}=L_{\rm SF}/N_{O*}$ together with
the ``single scattering approximation'' ($\tau=1$) of MQT, we may write
\be
v^2\sim \frac{L_{\rm SF}\,G}{2\,c\,f_g\,\sigma^2}\left(\frac{\Delta x/2}
{R}\right).
\ee
For $L_{\rm SF}=L_M$, where $L_M=4f_g\,c\,\sigma^4/G$ is maximum stellar 
luminosity derived in MQT, we find that
\be
v\sim\sigma\sqrt{\frac{\Delta x}{R}}\sim \sigma N^{-1/6}_{O*}.
\ee 
For a bright starbust galaxy with $N_{O*}\sim 10^6$, the radiation force
may lead to random fluid motions in the gas with characteristic velocities 
of only $\sim 1/10$ that of the isothermal velocity dispersion.  Therefore, 
it seems highly unlikely that radiation-driven feedback can lead to the 
unbinding of a galaxy's gaseous component.  The clustering of massive 
stars won't alleviate this major concern. 
Even if O stars are embedded together 100 at a time behind
a common dust sublimation layer, the 
random velocity $v$ increases by only a factor of $\sim 2$.

\section{Cosmic Ray Transfer with Destruction:  A Random Walk
Model for the Cosmic Ray Pressure Gradient Force}
\label{a: yukawa}

We begin with the equations relating CR pressure and flux for a
plane-parallel disk-like geometry.  First we assume that CR flux
$F_{\rm CR}$ is determined by
\be
\frac{d F_{_{\rm CR}}}{dz}=-t_{\pi}^{-1} 3 P_{_{\rm CR}} + Q_{_{\rm CR}}.
\label{e:mom0}
\ee
Here we have assumed $\gamma_{\rm CR}=4/3$, $t_{\pi}$ is the CR
lifetime to pion production (proton-proton scattering), $P_{_{\rm CR}}$
is CR pressure, and $Q_{_{\rm CR}}$ is the CR source term.  Other
processes may destroy CRs, but we expect pion production to
be most important for starforming galaxies.  In the diffusion
dominated limit $F_{_{\rm CR}}$ has a simple relation to $P_{_{\rm CR}}$ 
\be
\frac{d P_{_{\rm CR}}}{dz}=-\frac{F_{_{\rm CR}}}{c\,\lambda_{_{\rm CR}}}.
\label{e:mom1}
\ee

Combining eqs. (\ref{e:mom0}) and (\ref{e:mom1}), and using the
definition $\lambda_{\pi} \equiv t_{\pi} c$ we find 
\be \frac{d^2
P_{_{\rm CR}}}{dz^2}=\frac{3 P_{_{\rm CR}} }{\lambda_{_{\rm CR}}
\lambda_{\pi}}-\frac{Q_{_{\rm CR}}}{\lambda_{_{\rm CR}} c}.  \nonumber
\ee 
This equation now resembles the angle averaged radiative transfer
equation assuming the Eddington relation $P_{_{\rm CR}}=E_{_{\rm
CR}}/3$ applies.  Following the normal prescription for a scattering
dominated radiative atmosphere (see e.g. Rybicki \& Lightman 1979), we
now define an effective mean free path $l_{_{\rm CR}} \equiv
(\lambda_{_{\rm CR}} \lambda_{\pi}/3)^{1/2}$ and $d \tau_*
\equiv - l_{_{\rm CR}}^{-1} dz$.  This yields the familiar form
\be
\frac{d^2 P_{_{\rm CR}}}{d \tau_*^2}=P_{_{\rm CR}} - 
\frac{\lambda_{\pi} Q_{_{\rm CR}}}{3 c}.
\label{e:trans}
\ee

For simplicity we will assume $\lambda_{\pi} Q_{_{\rm CR}}$ is
constant.  With this assumption $Q_{_{\rm CR}}=F_{\rm max}/H$ where $H$
is the disk scale height and $F_{\rm max} \simeq L_{_{\rm CR}}/(4\pi
R^2)$.  The solution to eq. (\ref{e:trans}) is then given by
\be
P_{_{\rm CR}}=P_- e^{-\tau_*} + P_+ e^{\tau_*} 
+\frac{\lambda_{\pi} Q_{_{\rm CR}}}{3 c}.
\nonumber
\ee
In order to specify $P_+$ and $P_-$, we need two boundary conditions.
First, we will assume that disk has a midplane defined by $\tau_{\rm
m} = H/\lambda_{_{\rm CR}}$ where $F_{_{\rm CR}}$ vanishes due to
symmetry.  At the surface, we assume there is no incoming CRs.  In the
two-stream approximation (see e.g. Rybicki \& Lightman 1979) this
gives
\be
\frac{1}{\sqrt{3}} \frac{d P_{_{\rm CR}}}{d \tau_{_{\rm CR}}}=P_{_{\rm CR}},
\nonumber
\ee
at $\tau_{_{\rm CR}}=0$.
With these assumptions, it is easy to show that
\be
P_{_{\rm CR}} = \frac{\lambda_{\pi} Q_{_{\rm CR}}}{3 c} \left[1-
\frac{exp{(-\tau_*)}+exp{(\tau_*-2H/l_{_{\rm CR}})}}
{1+\sqrt{\epsilon}+exp{(-2H/l_{_{\rm CR}})}(1-\sqrt{\epsilon}) } \right],
\label{e:pcr}
\ee
where we have defined $\epsilon \equiv \lambda_{_{\rm
CR}}/\lambda_{\pi}$ (not to be confused with supernovae efficiency)
for convenience.  Note that, by assumption, $\epsilon \ll 1$.

A number of results are immediately apparent from
eq. (\ref{e:pcr}). First, we confirm that $P_{_{\rm CR}}$ is a maximum
at the midplane.  Furthermore, in the limit $H \gg l_{_{\rm CR}}$
where $P_{_{\rm CR}}=t_\pi Q_{_{\rm CR}}/3$.  This simply shows that
if all CR are destroyed before they escape the total energy density of
CRs is equal to the rate at which they are produced multiplied by the
time it takes to destroy them.  Note that this is the assumption of
Jubelgas et al. (2006).

\begin{figure}
\begin{center}
\includegraphics{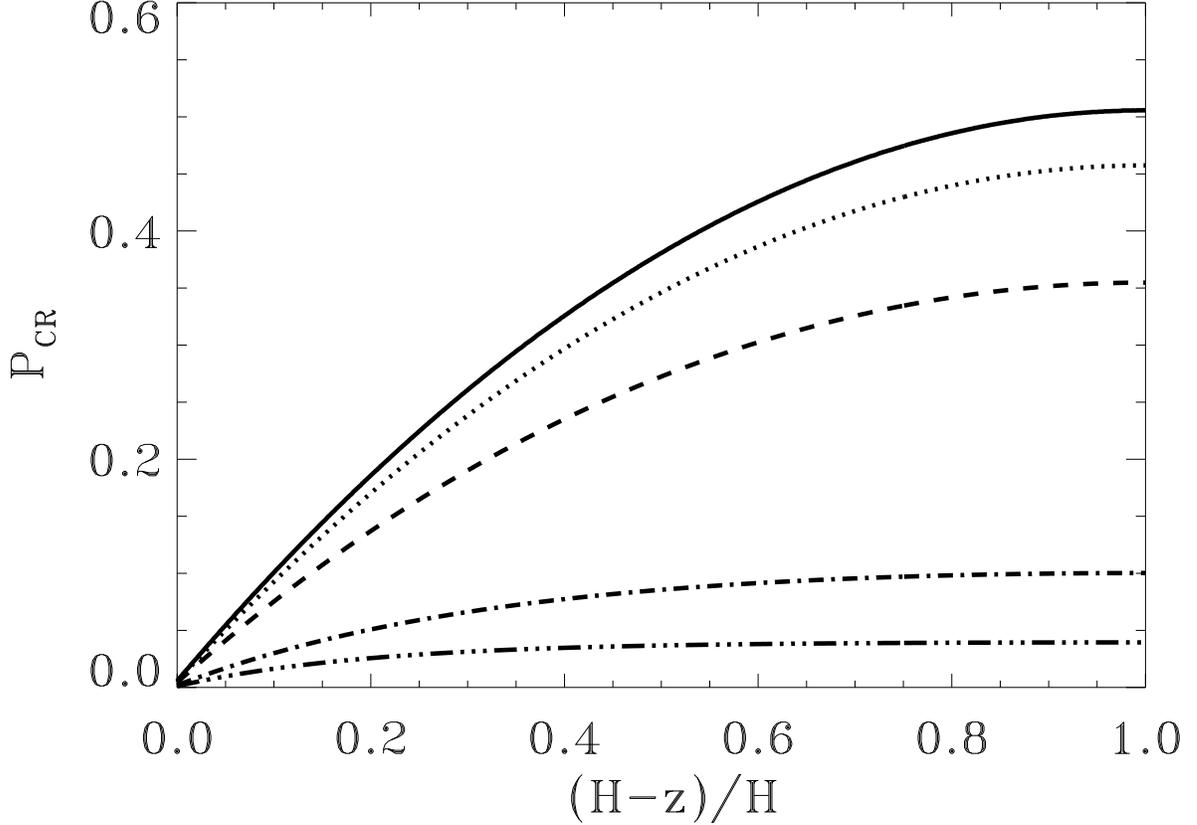}
\caption{Cosmic ray pressure as a function of atmospheric depth 
for varying levels of cosmic ray destruction.  The midplane corresponds
to $z=0$.  The vertical axis is the cosmic ray 
pressure in units of $H\, F_{\rm max}/{c\lambda_{_{\rm CR}}}$.
The ratio of the cosmic $\lambda_{_{\rm CR}}/H=0.1$.
The solid line corresponds to pure scattering and no 
destruction i.e., $H/l_{_{\rm CR}}=0$, while the 
$H/l_{_{\rm CR}}=0.5$ (dotted), 1.0 (dashed), 3.0 (dot-dashed)
and 5.0 (triple-dot dashed).  For relatively large levels
of CR destruction, the midplane pressure is reduced by a factor
$\sim l^2_{_{\rm CR}}/H^2$, as in the case of Jubelgas et al. (2006).}
\label{f: pressure}
\end{center}
\end{figure}

\begin{figure}
\begin{center}
\includegraphics{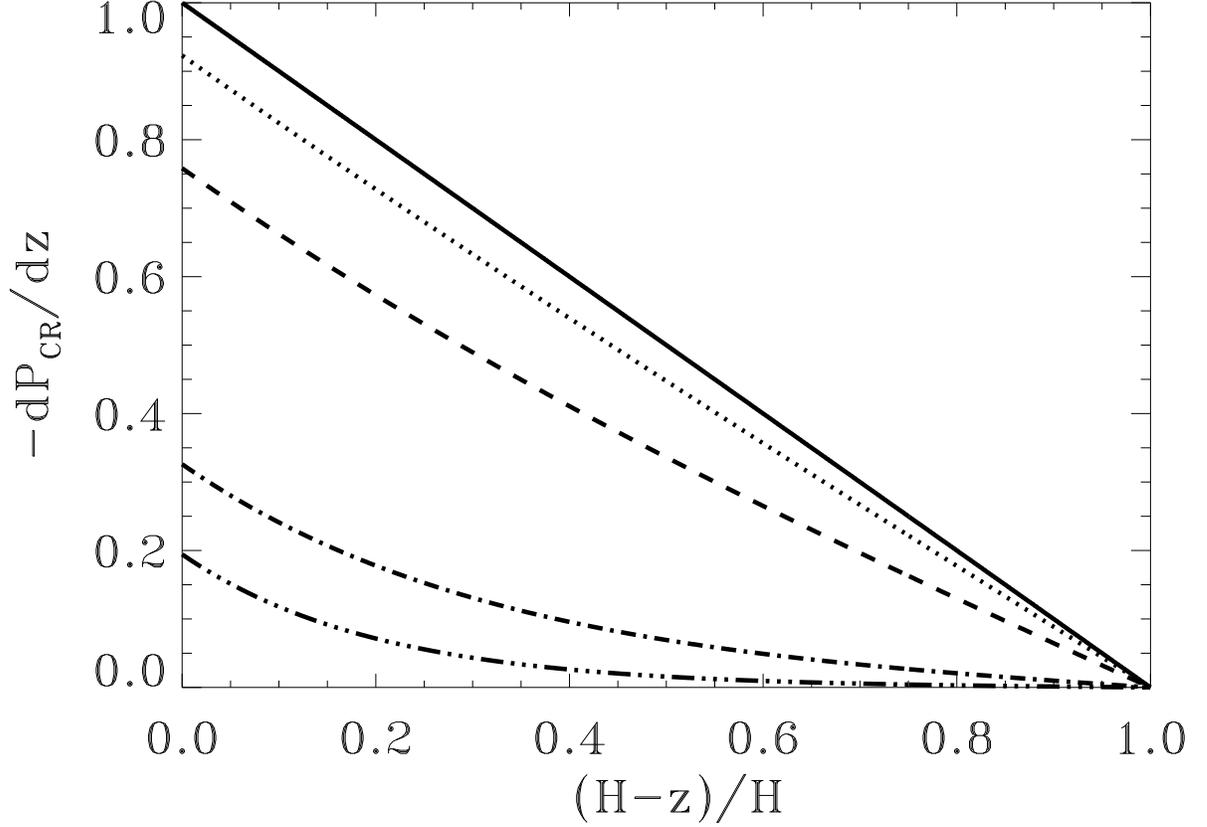}
\caption{Cosmic ray pressure gradient
as a function of atmospheric depth 
for varying levels of cosmic ray destruction.  The values for 
all of the input parameters are identical to those chosen in 
Figure \ref{f: pressure}.  The pressure gradient and 
therefore, the vertical cosmic ray pressure gradient force 
vanishes at the midplane in all cases, due to symmetry.  Furthermore,
the cosmic ray force and pressure gradient attain their maximum values
at the top of the atmosphere, whether or not destruction is 
important.  For the cases in which CR destruction is important, 
the maximum value for the CR force is reduced by 
$l_{_{\rm CR}}/H$.  It follows that the  Eddington limit in 
cosmic rays $L_{_{\rm edd, CR}}$ increases by a factor of $H/l_{_{\rm CR}}$
relative to the pure scattering case, for situation where 
CR destruction is important.}
\label{f: pressure_grad}
\end{center}
\end{figure}

Ultimately, we are interested in the force (per unit volume) the CRs provide, 
which is given by
\be
\frac{d P_{_{\rm CR}}}{dz}=\frac{\lambda_{\pi} Q_{_{\rm CR}}}{3 c l_{_{\rm CR}}}
\frac{exp{(-\tau_*)}-exp{(\tau_*-2H/l_{_{\rm CR}})}}
{1+\sqrt{\epsilon}+exp{(-2H/l_{_{\rm CR}})}(1-\sqrt{\epsilon})}.
\label{e:force}
\ee
It is clear from the form of the numerator that this force vanishes at the
midplane.  This must be the case, since the CR flux vanishes at the midplane
due to the assumed symmetry.

Now consider the force near the surface.  We'll first examine eq. 
(\ref{e:force}) in the limit that $H \ll l_{_{\rm CR}}$. In this case the 
numerator and denominator simplify considerably and we find
\be
\frac{d P_{_{\rm CR}}}{dz} \simeq \frac{\lambda_{\pi} Q_{_{\rm CR}}}
{3 c l_{_{\rm CR}}^2}
z.
\nonumber
\ee
For this case we see that the force increases linearly with height $z$ above the
midplane.  The maximum force is reached at the surface $z \sim H$ where
\be
\frac{d P_{_{\rm CR}}}{dz} \simeq \frac{\lambda_\pi Q_{_{\rm CR}} H}
{3 l_{_{\rm CR}}^2 c}
\simeq \frac{Q_{_{\rm CR}} H}{\lambda_{_{\rm CR}} c}
\simeq \frac{F_{\rm max}}{\lambda_{_{\rm CR}} c},
\nonumber
\ee
which is the expected result if CR destruction is negligible.

Now consider eq. (\ref{e:force}) in the limit that $H \gtrsim l_{_{\rm CR}}$.
Since we are primarily interested in the force near the surface, we will
ignore the $exp{(-2H/l_{_{\rm CR}})}$ dependent terms since they only provide
order unity corrections. We then find
\be
\frac{d P_{_{\rm CR}}}{dz} \simeq \frac{\lambda_{\pi} Q_{_{\rm CR}}}
{3 c l_{_{\rm CR}}} \frac{exp{(-\tau_*)}}{1+\sqrt{\epsilon}}.
\ee
Dropping the $\sqrt{\epsilon}$ term, which is small, and using the above 
definition for $Q_{_{\rm CR}}$, this simplifies to
\be
\frac{d P_{_{\rm CR}}}{dz} \simeq \frac{l_{_{\rm CR}}}{H}
\frac{F_{\rm max}}{\lambda_{_{\rm CR}} c} exp{(-\tau_*)}.
\nonumber
\ee
We can see that force is reduce roughly exponentially with
$\tau_*$ as we move deeper in the atmosphere, eventually
going to zero at the midplane.  The maximum force is at the surface
and is equal to
\be
\frac{d P_{_{\rm CR}}}{dz} \simeq \frac{l_{_{\rm CR}}}{H}
\frac{F_{\rm max}}{\lambda_{_{\rm CR}} c} \propto \sqrt{t_\pi}.
\nonumber
\ee
With the help of figures \ref{f: pressure} and \ref{f: pressure_grad},
the above expression has a clear interpretation.  
The force exerted by the CRs is
still given by $F_{_{\rm CR}}/(\lambda_{_{\rm CR}} c)$, but $F_{_{\rm
CR}}$ is reduced from the maximum flux $F_{\rm max}$ by a factor of
$l_{_{\rm CR}}/H$ because only CRs produced within $\sim l_{_{\rm
CR}}$ of the surface can contribute to $F_{_{\rm CR}}$ at the surface.
As we move deeper within the atmosphere the force is reduced because
downward moving CR from above increasingly offset the upward moving
CRs from below.

\section{Appendix D: Momentum and Energy Conservation for 
Optically Thick Radiation-Driven Winds}
\label{a:wind}

Here, CRs are taken to provide the radiation pressure that 
ultimately drives out a given galaxy's gaseous 
component.  By making the simplifying assumption that the force
exerted by the CRs upon the gas greatly exceeds the gravitational and
thermal pressure forces, the luminosity in CRs may be related to 
the outgoing momentum of the wind by
\be
\dot{P}_{w}=\dot{M}v_w=\frac{\Lcr}{c}\tau,
\label{a:wind1}
\ee
where $P_w$, $\dot{M}$, and $v_w$ is the momentum, mass loss rate,
and terminal velocity of the wind, respectively 
(Lamers \& Cassinelli 1999).  In eq. (\ref{a:wind1}), the 
optical depth of the wind $\tau$ cannot be arbitrarily large and is
constrained by energy conservation of the flow i.e.,
\be
\Lcr\leq \frac{1}{2}\dot{M}v^2_w.
\ee     
It follows that $\dot{M}v_w\leq 2\Lcr/v_w$, which gives a upper 
limit for the optical depth
\be
\tau\leq\frac{2c}{v_w}.
\ee
In the extreme limit, where $\tau=2c/v_w$, CRs are completely redshifted away
due to adiabatic expansion since they have transferred all of their 
energy to the matter. 

\section{Appendix E: Turbulent Power of Molecular Clouds 
Regulated by Cosmic Rays}
\label{a:turb}

The observation that the light to mass ratio ${\mathcal R}$ of both
``fully populated'' molecular clouds and bright dense regions of
starbursting galaxies are equal imply the presence of a common
feedback mechanism.  Here, we explore the possibility of whether or
not the production and transport of CRs {\it within} giant 
molecular clouds (GMCs) can regulate star formation.

In the Galaxy, there is $\sim 10^9M_{\odot}$ of molecular gas
responsible for the formation of stars and most of this mass is in the
form of GMCs (Binney \& Tremaine 1987).  We work under the assumption
that turbulence is the dominant form of pressure support and that the
molecular cloud is in virial balance.

The amount of power, or luminosity, required to maintain steady 
turbulence in a GMC, $L_{\rm turb}$, may be expressed
in terms of $M_c$, $v_{l_c}$, and $\Delta t_{\rm turb}$, which 
represent the mass of the GMC, the turbulent velocity of the 
outer scale $l_c$, and the eddy turnover time on the outer 
scale, respectively.  We have
\be
L_{\rm turb}\simeq M_{ c}\frac{v^2_{l_c}}{\Delta t_{\rm turb}}\simeq 
M_c\frac{v^3_{l_c}}{l_c}\simeq 2\times 10^{36} M_6v^3_3l^{-1}_{10}
\,{\rm erg\,s^{-1}},
\ee      
where $\Delta t_{\rm turb}\simeq l_c/v_{l_c}$ and $M_6$, 
$v_3$ and $l_{10}$ is $M_c$ in units of $10^6M_{\odot}$, 
$l_c$ in units of 10 pc,  $v_{l_c}$ in units of $3\,{\rm km s^{-1}}$, 
respectively.  

Now, we estimate an upper limit for the amount of power in 
CRs, which may be injected into the GMC.  As usual, the source
of CR power is assumed to be core-collapse SNe and well as 
winds from massive stars.  Again, we take into account that only $\sim 10
-20\%$ of massive
stars are embedded i.e., in physical contact with their natal GMCs,
at any given moment  (Wood \& Churchwell 1989).
From this, we infer that a similar fraction of the SNe 
and stellar wind power comes into contact with its ancestral 
GMC.  Putting this together, the CR energy injection rate
that a GMC is subject reads
\be
L^{\rm cloud}_{\rm CR}\simeq \xi\frac{M_c}{M_{\rm mol}}\Lcr
\simeq 6\times 10^{36}\xi_{0.1}M_6M^{-1}_{\rm mol,9}\epsilon_{6}
\dot{m}_{\rm _{SF}}\,{\rm erg\,s^{-1}}.
\ee
Here $\xi$ and $M_{\rm mol}$ represent the fraction of massive stars 
that reside within their parent GMC at any given time 
and the entire molecular mass 
content of the Galaxy and furthermore, $M_{\rm mol,9}$ and $\xi_{0.1}$
are $M_{\rm mol}$ and $\xi$ in units of $10^9M_{\odot}$ and 0.1, 
respectively.

Understanding the parameter $\xi$ deserves some attention.
One interpertation is that the fraction $\xi$ results from the 
possibility that every massive star eventually leaves its 
molecular cloud of origin before it expires into a supernova.
The other interpretation is that a fraction $\xi$ of massive 
stars remain embedded for their entire lifetime.  With respect 
to the former, CR feedback in GMCs plays a role only if stellar
winds from massive stars effectively produce CR protons.  For the 
latter, CR feedback may be important if either core-collapse
SNe or stellar are responsible for their production.  Note that
$\xi$ is likely larger in the latter case.  

On energetic grounds, the above expressions indicate that CRs indeed
may be a crucial agent of self-regulation for GMCs.

\end{document}